\documentclass[%
 aip,
rsi,%
 amsmath,amssymb,
 reprint,%
]{revtex4-1}
\usepackage{amsmath,amssymb,amstext,amsthm}
\usepackage[pdftex]{graphicx}
\usepackage{braket}
\usepackage{bbm}
\usepackage{xcolor}
\definecolor{dred}{rgb}{.8,0.2,.2}
\definecolor{ddred}{rgb}{.8,0.5,.5}
\definecolor{dblue}{rgb}{.2,0.2,.8}
\usepackage[pdftex,pagebackref=false]{hyperref}
\hypersetup{
pdftitle={Noise Refocusing in a Double Loop Mach-Zehnder Neutron Interferometer},
pdfauthor={J. Nsofini},
pdfsubject={Quantum information, quantum physics, neutron interferometry, quantum coherence}
	pdfnewwindow=true, 
	colorlinks=true, 
	linkcolor=blue, 
	citecolor=blue, 
	filecolor=blue, 
	urlcolor=blue
}

\def\beq{\begin{equation}}
\def\eeq{\end{equation}}
\def\bsp{\begin{split}}
\def\esp{\end{split}}
\def\bea{\begin{eqnarray}}
\def\eea{\end{eqnarray}}
\def\ba{\begin{array}}
\def\ea{\end{array}}

\graphicspath{{Figures/}}
\begin{document}

%:Title Info
\title{Noise Refocusing in a Five-blade Neutron Interferometer}
%%%Commenting every one's name
%% Joachim
\author{J. Nsofini} 
\email{jnsofini@uwaterloo.ca}
\affiliation{Department of Physics, University of Waterloo, Waterloo, ON, Canada, N2L3G1}
\affiliation{Institute for Quantum Computing, University of Waterloo,  Waterloo, ON, Canada, N2L3G1}
%%Dusan
\author{D. Sarenac}
\affiliation{Department of Physics, University of Waterloo, Waterloo, ON, Canada, N2L3G1}
\affiliation{Institute for Quantum Computing, University of Waterloo,  Waterloo, ON, Canada, N2L3G1} 
%%Kamyar
\author{K. Ghofrani}
\affiliation{Institute for Quantum Computing, University of Waterloo,  Waterloo, ON, Canada, N2L3G1} 
\affiliation{Department of Chemistry, University of Waterloo, Waterloo, ON, Canada, N2L3G1}
%%Mike
\author{M. G. Huber}
\affiliation{National Institute of Standards and Technology, Gaithersburg, MD 20899, USA}
%%Arif
\author{M. Arif}
\affiliation{National Institute of Standards and Technology, Gaithersburg, MD 20899, USA}
%% David
\author{D. G. Cory}
\affiliation{Institute for Quantum Computing, University of Waterloo,  Waterloo, ON, Canada, N2L3G1} 
\affiliation{Department of Chemistry, University of Waterloo, Waterloo, ON, Canada, N2L3G1}
\affiliation{Perimeter Institute for Theoretical Physics, Waterloo, ON, Canada, N2L2Y5}
\affiliation{Canadian Institute for Advanced Research, Toronto, Ontario, Canada, M5G 1Z8}
%% Dmitry
\author{D. A. Pushin}
\affiliation{Department of Physics, University of Waterloo, Waterloo, ON, Canada, N2L3G1}
\affiliation{Institute for Quantum Computing, University of Waterloo,  Waterloo, ON, Canada, N2L3G1}

\begin{abstract}
We provide a quantum information description of a proposed five-blade  neutron interferometer geometry and show that it is robust against low frequency mechanical vibrations and dephasing due to the dynamical phase. The extent to which the dynamical phase affects the contrast in a neutron interferometer is experimentally shown. In our model, we consider the coherent evolution of a neutron wavepacket in an interferometer crystal blade and simulate the effect of mechanical vibrations and momentum spread of the neutron through the interferometer. The standard three-blade neutron interferometer is shown to be immune to dynamical phase noise but prone to noise from mechanical vibrations, the decoherence free subspace four-blade neutron interferometer is shown to be immune to mechanical vibration noise but prone to noise from the dynamical phase, while the proposed five-blade neutron interferometer is shown to be immune to both low-frequency mechanical vibration noise and dynamical phase noise.
\end{abstract}

%\pacs{}
\maketitle

%==============================================================
% Introduction
%==============================================================
 \section{Introduction}

%maybe a part about uses of nutron interfrometers? 
%maybe just stick to NI and not mention atomic interferometry. theres just no space for such a general intro
Matter wave interferometry is a powerful and extremely sensitive tool to probe effects ranging from material properties to foundational physics \cite{ Rauch_2000_Book,Cronin_2009_RevModPhys,li2016neutron,sarenac2016holography}. Although high sensitivity and accuracy are achieved due to matter waves' small deBroglie wavelength and statistical inference, these massive particles couple to external degrees of freedom (DOF) leading to loss of coherence. 
Loss of coherence as a result of non-refocused phases has been a subject of study in matter wave interferometry \cite{Rauch_2000_Book,Sudo_2006_Book, Hellweg_2003_PRL,Marquardt_2004_PRB}. In this work, we specifically discuss concepts applied to a neutron interferometer (NI).  However, they could easily applied to other matter wave interferometers.
%and therefore most of the concept presented here for a neutron interferometer (NI) could be applied to other examples of matter wave interferometry.
Isolation and control techniques have been developed to deal with some classes of noise \cite{Arif_1994_VibrMonitCont,Shahi_2016_NIM, Pushin_2015_AHEP,Saggu_RevSciInstrum_2016} but low-frequency vibrational noise still persists in these setups. The quest for noise-immune neutron interferometry motivated the design of the four-blade neutron interferometer with a decoherence free subspace (DFS) \cite{Pushin_2009_PRA,Pushin_2011_PRL} which is robust to noise originating from mechanical vibration. 
Although the four-blade NI is robust against low-frequency vibrations, we will show that it is prone to dynamical phase noise. 
%Dynamical diffraction phases \cite{Lemmel_2013_ActaCrys} which result from variations in a blade's transmission amplitude  limit the coherence in a neutron interferometer and therefore should be refocused. 

During dynamical diffraction (DD) from a perfect crystal, a phase shift is introduced due to diffraction in the vicinity of the Bragg condition \cite{Lemmel_2007_PRB, Springer_ActaCrysA,Zawisky_2011_NIM,Lemmel_2013_ActaCrys}. The so called \textit{dynamical phase} has tremendous
angular sensitivity, which a recent experiment has measured to be about 30$\pi$ rads per arsec deviation from the Bragg angle in a silicon [220]  crystal \cite{Potocar_2015_ActaCrysA}. This sensitivity may offer a possibility of extracting fundamental quantities such as the neutron-electron scattering length, short-range gravitational interactions, and the Debye-Waller factor \cite{Ioffe_2002_ApplPhysA, Wiedtfield_2006_PhysB,Green_2007_PRC}. 

The presence of the dynamical phase can lead to a reduction in the interferometry fringe visibility via a loss of coherence from a phase variation across the neutron beam \cite{Rauch_2000_Book,Lemmel_2010_PRA, Petrascheck_1976_PhysStatSol, Pushin_2008_PRL}. 
As a result, it is desirable  to remove the dynamical phase gradients. 
Such phases are naturally refocused in a three-blade NI but not the four-blade NI. Here,  we propose a five-blade NI geometry that is robust to dynamical phase noise and also refocuses low frequency mechanical vibrational noise like the four blade DFS NI.

This article is structured as follows: Section \ref{Sec:NIgeometries} gives a brief overview of the NI geometries considered including the proposed five-blade NI. In section \ref{Sec:DDEffect} we give an analysis of the effect of the dynamical phase on the three-blade, four-blade and five-blade neutron interferometers. 
The effect of external vibrations on each of the interferometer geometries is presented in section \ref{Sec:MechVib},
including a description of the noise in terms of the coherence function \cite{Petrascheck_1986_PRB,Rauch_1996_PRA, Rauch_2000_Book,Ewolf_2007_Book} to demonstrate the robustness of the five blade geometry to noise.

%==============================================================
% Geometries of the NI
%==============================================================

\section{Perfect Crystal Neutron Interferometers} 
%and Contrast Measurements}
\label{Sec:NIgeometries}

\begin{figure}
\center
\includegraphics[width=0.95\columnwidth]{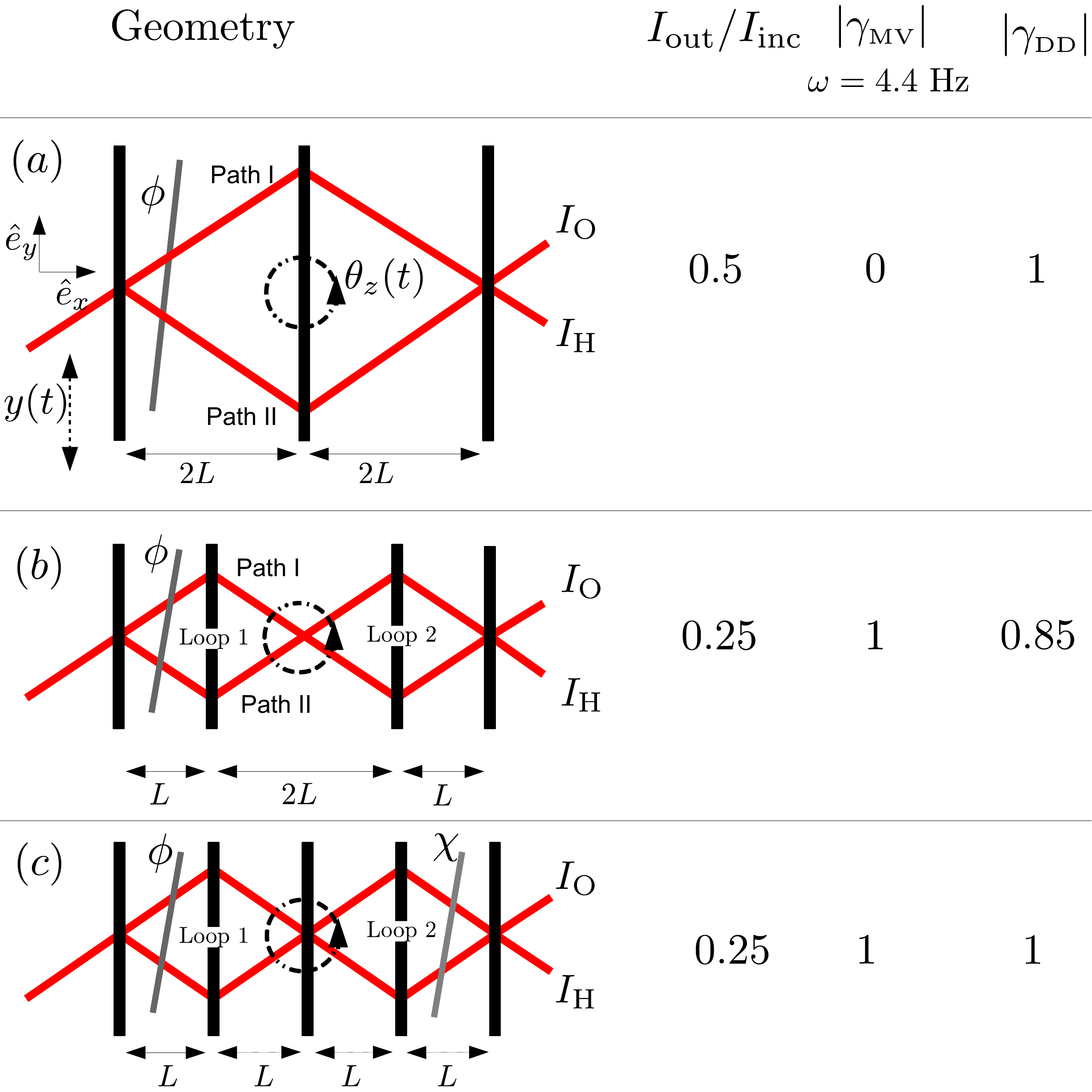} 
\caption{Sketch of different NI geometries with phase flags producing phases $\phi$ and $\chi$. 
The two detectors are the O-detector ($I_\text{O}$) and the H-detector ($I_\text{H}$). 
a) A symmetric three blade NI with phase flag $\phi$ and inter-blade distance 2$L$.
 b) Four blade (DFS) NI with inter-blade distances: $L$, 2$L$, $L$. c) Five blade (double MZ loop) NI with with inter-blade distances $L$.  The noise along the y-axis $y(t)$, and along around the z-axis $\theta_z(t)$, are modeled as sinusoidal. $I_\text{out}/I_{inc}$ is the ratio of the neutrons at the output $I_\text{out}=I_\text{O}+I_\text{H}$ to those at incidence $I_{inc}$, and where we have assumed $50:50$ beam splitters.  $|\gamma_\text{\tiny MV}|$ and $|\gamma_\text{\tiny DD}|$  are the absolute values of the coherence function with z-noise ($\omega=4.4$ Hz, see Fig.~\ref{Fig:ZNoise}) and dynamical phase noise simulated for the DFS interferometer dimensions of \cite{Pushin_2011_PRL}, respectively. In an ideal case these should be 1. }\label{Fig:345NI}
\end{figure}

A common NI geometry is the symmetric Laue-type which is machined from a perfect single crystal ingot of silicon and composed of several identical separate blades. A neutron incident on a blade in the NI is Bragg diffracted into two coherent beam paths. 
In this paper we adopt the quantum operator formalism of Bragg diffraction from a perfect crystal \cite{Nsofini_2016_PhysRevA}. 
	The path degree of freedom (DOF) is a two-level system that is defined by the sign of the momentum in the y-direction (see Fig.~\ref{Fig:345NI} for coordinate system); the path with $+k_y$ is labelled as path I and the path with $-k_y$ is labelled as path II. This two level system is isomorphic to a Bloch sphere \cite{Ramsey_1950_PhysRev}. %Rotations are then defined by the Pauli matrices.

%By adding a phase flag and rotating inside any neutron interferometer a  phase difference is introduced between the two paths resulting to an intensity oscillation between the two detectors at the output.
%===========================================
%: Fig: Neutron Interferometers
%=================================================================================================================
%=================================================================================================================

The three-blade NI considered (see Fig.~\ref{Fig:345NI}a) consists of three blades separated by the same distance ($2L$) in the Laue geometry. The second blade redirects the two paths to the third blade where they recombine and interfere. Each of the blades of the NI acts as a beam splitter.
However, due to post selection on only those neutrons that reach the detector, the second blade actually implements a perfect $\pi$ pulse. This enables a very simple and robust picture of the physics.

%Although there is a finite probability of a neutron being transmitted at the second blade and escaping the NI, we can ignore these neutrons by postselecting on the neutrons that arrive at the detectors. Hence by simple renormalizing we ignore these transmitted neutrons and treat the second blade as a perfect mirror \cite{Pushin_2007_Thesis}. 

In the four-blade NI (Fig.~\ref{Fig:345NI}b), the situation is similar to the three blade NI with the difference that the two paths are redirected twice (with no mixing of states in the center of the interferometer) before reaching the last blade. This four-blade NI posses a DFS for low-frequency mechanical vibrational noise  which significantly affects the three blade NI \cite{Pushin_2009_PRA, Pushin_2011_PRL}.  
%In this paper we explore classes of noise that affect the DFS interferometer (dynamical diffraction noise, external vibrations, magnetic) and then propose a five blade NI which is robust to this class of noise.

The proposed five-blade NI (Fig.~\ref{Fig:345NI}c) can be thought of as two coupled Mach-Zehnder NIs. It is similar to the four-blade NI in that the neutrons are redirected twice but differs since the neutrons  interfere on the additional blade in the middle. 

One figure of merit quantifying the quality of the interferometry setup is  the fringe visibility or contrast. 
By introducing a phase difference between the two paths, the intensity at the exit oscillates between the intensities at the O-beam $(I_O)$ and H-beam $(I_H)$. 
From the intensity oscillations, the contrast is defined by
\beq
\mathcal{V}=\frac{{I}_\mathrm{max} -{I}_\mathrm{min}}{{I}_\mathrm{max} +{I}_\mathrm{min}},\label{Eq:Contrast}
\eeq
where, $I_\mathrm{max}$ and $I_\mathrm{min}$ are the maximum and minimum intensities. Contrast is related to the coherence in the path DOF in an NI. 
Coherence refers to the ability of the two paths to interfere. It has been extensively studied in matter wave and photon optics \cite{Rauch_1996_PRA,Ewolf_2007_Book,Glauber_1963_PhysRev}. 

%If an incident state $\ket{\Psi}$ evolves under the operator $U(\varphi)$ to $\ket{\Psi}_\varphi$, the coherence function $\Gamma$ and contrast $\mathcal{C}$ is defined as the overlap
%\begin{align}
%\Gamma(\varphi)=\bra{\Psi}\Psi\rangle_\varphi,\quad\mathcal{C}=\max_{\varphi}|\Gamma(\varphi)|-\min_{\varphi}|\Gamma(\varphi)|
%\end{align}
% Two quantities are used to quantify the quality of the interferometer setup; namely the  "Visibility (${\cal V}$)" and the "Contrast ($\Gamma$)". They are defined by: 
%\bea\label{coh}
%{\cal V}=\text{I}_\text{max}-\text{I}_\text{min},
% \quad \Gamma=\frac{\text{I}_\text{max}-\text{I}_\text{min}}%{\text{I}_\text{max}+\text{I}_\text{min}}
%\eea
%where $\text{I}_\text{max}$ and $\text{I}_\text{min}$ correspond to the maximum and minimum intensities at the detector.\todo{COME BACK} 

%================================================================================================
% Dynamical effect of Blades
%==========================================================================================
\section{Effect of dynamical phase}
\label{Sec:DDEffect}
The beam splitting in each of the blades of the NI is govern by the theory of dynamical diffraction. The theory of DD describes the interaction of matter waves and x-rays with a perfect crystal lattice when incident at Bragg condition \cite{Zachariasen_1945_Book,Rauch_1978_TopCurPhys, Sears_1989_Book,Bonse_1977_Book,  Batterman_1964_RevModPhys, Authier_2006_Book}.
Perfect crystals coherently split a neutron beam into two components with properties defined by the periodicity of the crystal lattice and the energy of the neutron \cite{Zeilinger_1981_AJP,Lemmel_2013_ActaCrys}. 
The mathematical formulation of the theory of DD is quite cumbersome and we have shown recently that we may use a simplified quantum information approach \cite{Nsofini_2016_PhysRevA}. 
Denoting the states corresponding to paths I and II as $\ket{\text{I}}$ and $\ket{\text{II}}$  and  the operator of the blade as $U_B$, the states after the first blade of an NI is 
\begin{align}
U_B\ket{\text{I}}&=t\ket{\text{I}}+r\ket{\text{II}},\ U_B\ket{\text{II}}=\overline{r}\ket{\text{I}}+\overline{t}\ket{\text{II}},
\end{align}
where the transmission and reflection coefficients  satisfy  $|t|^2+|r|^2=1$, and $\overline{r}=-r^*$, $\overline{t}=t^*$.

 Due to symmetry, the Bragg diffraction is required to take the same form if the crystal is rotated by 180$^\circ$. The crystal blade operator can be expressed as a composite sequence of rotations,
\bea
U_B=R_z(\phi_t)R_{xy}(\phi_r,\alpha)R_z(\phi_t)
\eea
with the standard definitions of Bloch sphere rotations
\bea
R_z(\phi_t)&=&\exp(i\phi_t\sigma_z/2),\\ R_{xy}(\phi_r,\alpha)&=&\exp(i\alpha(\cos(\phi_r)\sigma_x+\sin(\phi_r)\sigma_y)/2)
\eea
where  $\sigma_x=\ket{\text{I}}\bra{\text{II}}+\ket{\text{II}}\bra{\text{I}}$, $\sigma_y=-i\ket{\text{I}}\bra{\text{II}}+i\ket{\text{II}}\bra{\text{I}}$, $\sigma_z=\ket{\text{I}}\bra{\text{I}}-\ket{\text{II}}\bra{\text{II}}$ are the Pauli operators, $\phi_t=\arg[t]$, and $\phi_r=\arg[r]$.
By definition the dynamical phase is $\phi_t=\arg[t]$, while the phase between the two paths in an interferometer is $\beta=\phi_t-\phi_r$. Without loss of generality, we will limit the $R_{xy}$ rotation to be along $\sigma_x$, thereby effectively setting $\phi_r=0$. This is justified because $\phi_r$ is a small linear contribution.
This leads us to hypothesize a composite crystal blade operator \cite{Nsofini_2016_PhysRevA}
\bea
U_B=R_z(\beta)R_x(\alpha)R_z(\beta).
\eea
From these relations, one can identify the relation to the dynamical diffraction variables as
\bea
\beta=\arg[t], \ t=\cos(\alpha/2), \ \text{and}  \ r=\sin(\alpha/2),
\eea
with $\alpha\in[0,\pi]$. When  $\alpha=\pi/2$, the blade acts as a 50:50 beam splitter.

%\yo{moved to the intro}
%During DD from a perfect crystal, a phase shift occurs due to diffraction in the vicinity of the Bragg condition \cite{Lemmel_2007_PRB, Springer_ActaCrysA,Lemmel_2013_ActaCrys}. The presence of this phase has tremendousangular sensitivity, which a recent experiment has achieved about 30$\pi$ rads per arsec deviation from the Bragg angle of in a [220] crystal in the Laue geometry \cite{Potocar_2015_ActaCrysA}. This sensitivity may offer a possibility of extracting fundamental quantities including the neutron-electron scattering length, short-range gravitational interactions, the Debye-Waller factor \cite{Ioffe_2002_ApplPhysA, Wiedtfield_2006_PhysB,Green_2007_PRC}. In addition, the presence of the diffraction phase in an interferometer also reduces the interference patters. This is due to the fact that a loss of coherence and thus the contrast in neutron interferometry occurs when a phase gradient is present across the neutron beam \cite{Rauch_2000_Book,Lemmel_2010_PRA, Petrascheck_1976_PhysStatSol, Pushin_2008_PRL}. 
%As a result, it is a desirable process to remove any diffraction phase presence in any interferometry geometry. 

We will now apply the Bloch sphere rotation formalism described above to the three NI geometries to analyse the relevance of the dynamical phase in each case.

%====================================================================
\subsection{Three-blade Mach-Zehnder neutron interferometer}
 
In the three-blade NI the first and last blades each act as a composite rotation  $U_B=R_z(\beta)R_x(\alpha)R_z(\beta)$. The middle blade serves as a mirror to redirect the two paths onto the third blade, and hence is properly represented by $U_M=R_z(\beta)R_x(\pi)R_z(\beta)=R_x(\pi)$. 
With a phase difference $\phi$ (due to the phase flag) between path $\text{I}$ and path $\text{II}$ of the three-blade NI  (Fig.~\ref{Fig:345NI}a) the overall operation sequence is:
\begin{align}\nonumber
R_{3}&=U_B U_M R_z(\phi) U_B,\\\nonumber
&=R_z(\beta)R_x(\alpha)R_z(\beta)R_x(\pi)R_z(\phi)
R_z(\beta)R_x(\alpha)R_z(\beta),\\
&=R_z(\beta)R_x(\alpha)R_x(\pi)R_z(\phi) R_x(\alpha)R_z(\beta),\label{Eqn:R3}
\end{align}
where the identity  $R_x(\pi)=R_z(\beta)R_x(\pi)R_z(\beta)$ was used in the second line.
The first and last $R_z(\beta)$ rotations can be ignored  since the incoming beam is an eigenstate of $\sigma_z$ and the measurement is done along the $z$-basis. 
With an initial state $\ket{\text{I}}$, the intensity of neutrons at the O-detector and H-detector for $\alpha=\pi/2$ are,
\bea
I_{O3}(\phi)&=\frac{1}{2}(1+\cos\phi),\\
I_{H3}(\phi)&=\frac{1}{2}(1-\cos\phi).
\eea
The three-blade NI is therefore immune to dynamical noise as $\beta$ is refocused. It is worth noting that the resulting operation of the three blade NI is analogous to the Hahn echo sequence \cite{Hahn_1950_PhysRev}. 

\subsection{Four-blade neutron interferometer}
In the four-blade NI the operator of the first and fourth blades is $U_B$, and that of the second and third blades is $U_M$. With and initial state $\ket{\text{I}}$ and a phase difference $\phi$ between paths $\text{I}$ and $\text{II}$ (see Fig.~\ref{Fig:345NI}b) the overall operator sequence for the four blade NI is:
\begin{align}
\label{eqn:4DFsOperator}\nonumber
R_{4}&=U_B R_x(\pi) R_x(\pi) R_z(\phi) U_B,\\
%&=R_z(\beta)R_x(\alpha)R_z(\beta) R_x(\pi)
%\\\nonumber& R_x(\pi)R_z(\phi)R_z(\beta)R_x(\alpha)R_z(\beta),\\
%&=R_x(\alpha)R_z(\beta)R_x(\pi)R_x(\pi)R_z(\phi)R_x(\alpha),\\
&=R_z(\beta)R_x(\alpha)R_z(2\beta)R_z(\phi)R_x(\alpha)R_z(\beta),
\end{align}
The identity  $\mathbb{I}=R_x(\pi)_xR(\pi)$ was used in the second line. In the case where $\alpha=\pi/2$ the intensities at the O-beam and H-beam are given by
\bea\nonumber
I_{O4}(\phi)&%=|\bra{\text{I}}R_{4}\ket{\text{I}}|^2
=&\frac{1}{2}\Big[1-\cos(\phi+2\beta)\Big],\\\nonumber
I_{H4}(\phi)&%=|\bra{\text{II}}R_{4}\ket{\text{I}}|^2
=&\frac{1}{2}\Big[1+\cos(\phi+2\beta)\Big].
\eea
The presence of $\beta$ in the intensity implies that the dynamical phase is not refocused in the four blade NI. Upon averaging over neutrons with different momenta arriving at the detector, dephasing occurs in a four blade NI. 
The dephasing causes a reduction in the coherence and hence the contrast. The loss in contrast depends on the noise spectrum of $\beta$. The average neutron intensity at the detectors is
\bea\label{eqn:Ave4bphase}
\overline{I_{O4}}(\phi)&=&\frac{1}{2}\Big[1-\int d\beta p(\beta)\cos(\phi+2\beta)\Big],\\
\overline{I_{H4}}(\phi)&=&\frac{1}{2}\Big[1+\int d\beta p(\beta)\cos(\phi+2\beta)\Big],
\eea
where $p(\beta)$ is the probability density function.
%The effect of this rotation was mentioned in the early developments of neutron interferometry \cite{Ubonse,Petrascheck76} \todo{Discuss}.
The effect of this dynamical phase was pointed out in early works on neutron interferometry  \cite{Bonse_1979_Book,Petrascheck_1976_PhysStatSol}, but the extent to which it affects the coherence in a four-blade NI is not well quantified experimentally. The intensity can be re-written as
\bea\label{Eq:Obeam4DD}
\overline{I_{O4}}(\phi)&=&\frac{1}{2}\Big[1-|\gamma|\cos(\phi+\arg[\gamma])\Big],\\
\overline{I_{H4}}(\phi)&=&\frac{1}{2}\Big[1+|\gamma|\cos(\phi+\arg[\gamma])\Big],
\label{Eq:Hbeam4DD}
\eea
where,
\bea
\gamma=\int d\beta\ p(\beta) e^{i2\beta}
\eea
is the coherence function.
% Arbitrarily, the coherence function $\bra{\text{I}}R_{4}\ket{\text{I}}$ is becomes
% \bea
%\Gamma=\int \cos(\phi+2\beta) d\beta,
%\eea
%and the corresponding contrast is
% \bea
%\mathcal{C}&=&\max_\phi\int p(\beta)\cos(\phi+2\beta)^2d\beta.
%\eea
%The plot of contrast with standard deviation of the strength $\beta$ is give in Fig.~??. 
The presence of a phase distribution leads to a reduction in coherence and hence the contrast. This loss of contrast is usually small since the width of the distribution accepted by the NI crystal (Darwin width) is very narrow ($\sim 10^{-6}$ rad), thereby limiting the strength of the noise.

%=================================================================================================================
\begin{figure}
\center
\includegraphics[width=.9\columnwidth]{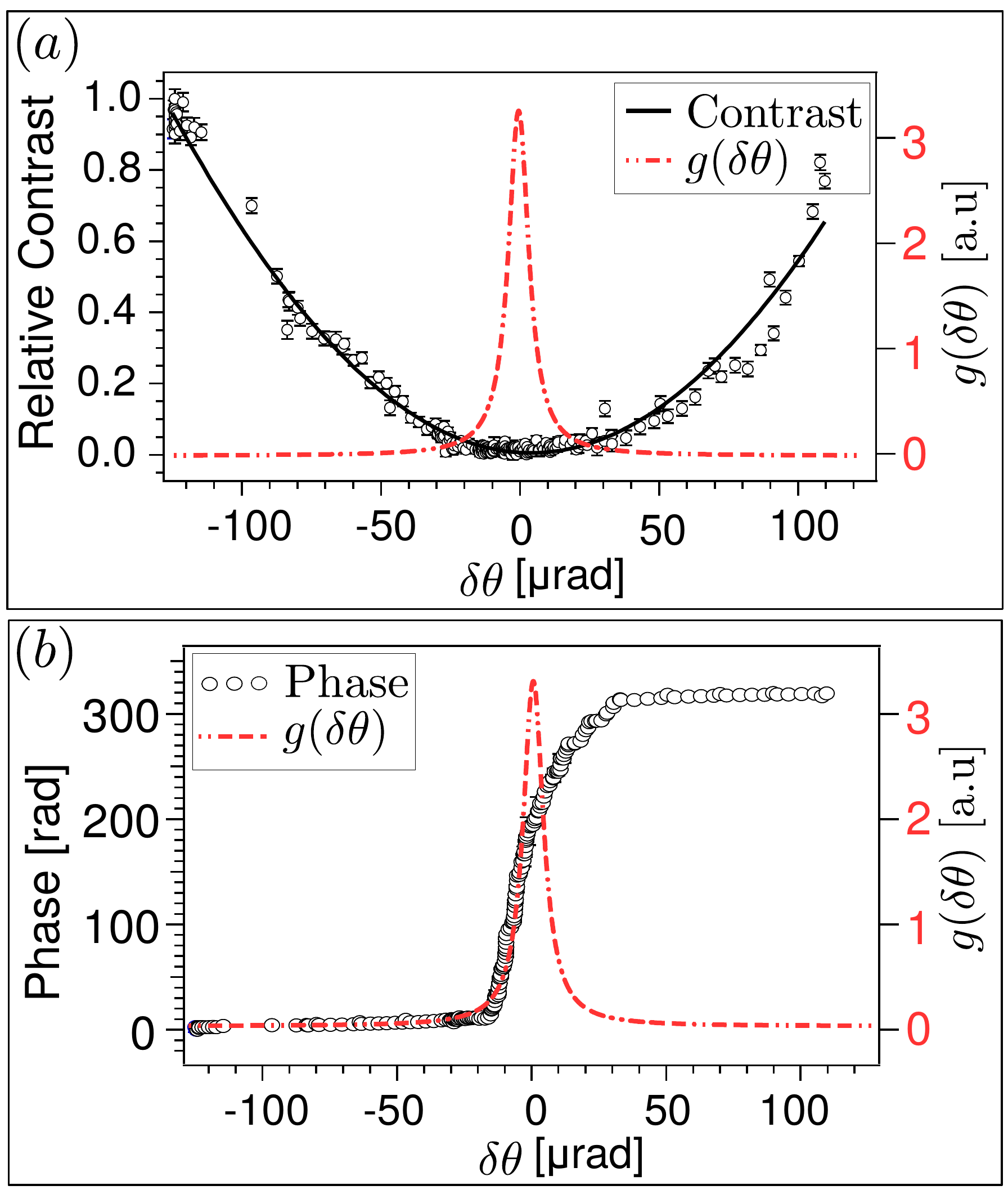} 
\caption{Measured effect of the dynamical phase on contrast \cite{Huber_Thesis_2008}. A 2 mm thick (111) Si crystal is added after the first blade of a three-blade neutron interferometer and rotated around the Bragg angle. (a) the contrast and momentum distribution $g(\delta\theta)$ plotted against the Si blade rotation $\delta\theta$. (b) the dynamical phase and the simulated momentum distribution $g(\delta\theta)$ plotted against the Si blade rotation $\delta\theta$.
 The full width at half maximum of the momentum distribution is equal to the Darwin width of the crystal  $\sigma_D=4.26\ \mu$rad.}\label{Fig:3to4BNI}
\end{figure}
%=================================================================================================================

\begin{figure}
\center
\includegraphics[width=0.95\columnwidth]{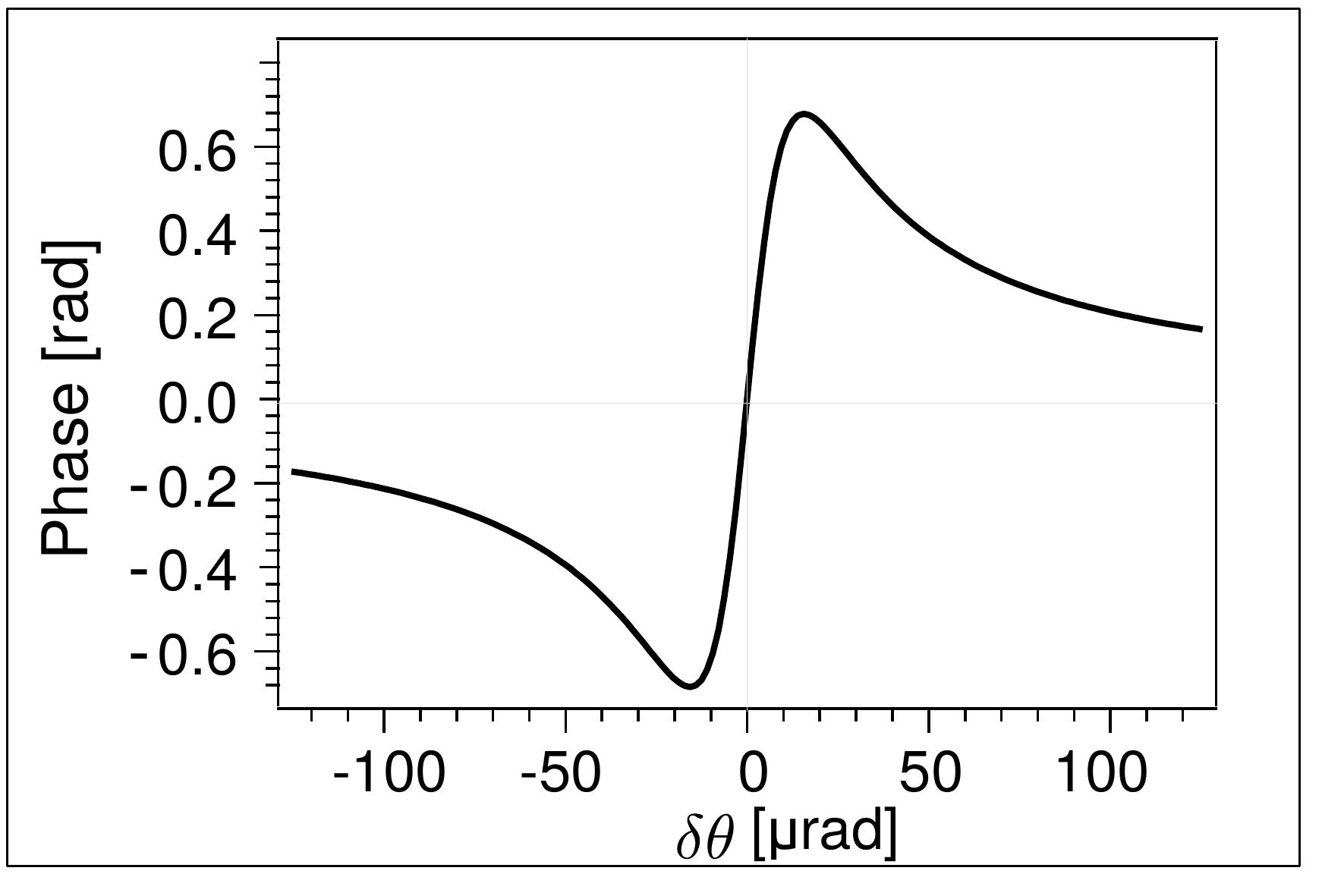} 
\caption{The dynamical phase separated from the total phase shown in Fig.~\ref{Fig:3to4BNI}b.  }\label{Fig:DDPhaseD}
\end{figure}

In an experiment to measure the neutron charge radius, the dynamical phase was measured as the extent to which the contrast is affected by the dynamical phase. In the experiment, a perfect Si crystal blade, of thickness 2 mm and crystallographic orientation [111], was added after the first blade of a three-blade NI \cite{Huber_Thesis_2008}. 
 When the crystal is aligned to the Bragg angle of the interferometer and the Bragg reflected beams are blocked, it replicates the dynamical phase that manifests itself in a four-blade NI. 
 Using $\alpha=\pi/2$ the normalized output intensity at the O-beam in this case is similar to Eq.~(\ref{Eq:Obeam4DD}), and can be expressed as
\bea
\overline{I_\textsc{{\tiny O}}}(\phi)&=\mathcal{A}_\text{\tiny O}-|\mathcal{B}_\textsc{{\tiny O}}|\cos(\phi+\arg[\mathcal{B}_\textsc{{\tiny O}}]),
\eea
where as shown in \cite{Lemmel_2013_ActaCrys},
\begin{align}
\mathcal{A}_\textsc{{\tiny O}}&=\!\int\! d\delta\theta  \ g(\delta\theta), \qquad
\mathcal{B}_\textsc{{\tiny O}}=\!\int\! d\delta\theta\  g(\delta\theta) e^{i\beta}, \\
 g(\delta\theta)&=\frac{\sigma_\theta/\pi}{\sigma_\theta^2+ (\delta\theta-\delta\theta_0)^2},
\end{align}
%\begin{align}
%\mathcal{A}_\textsc{{\tiny O}}&=\!\int\! d\delta\theta  \ p_\textsc{{\tiny O}}(\delta\theta), \qquad
%\mathcal{B}_\textsc{{\tiny O}}=\!\int\! d\delta\theta\  p_\textsc{{\tiny O}}(\delta\theta) e^{i\beta}, \\
% p_\textsc{{\tiny o}}(\delta\theta)&=\frac{\sigma_\theta/\pi}{\sigma_\theta^2+ (\delta\theta-\delta\theta)^2},
%\end{align}
%\begin{align*}
%\mathcal{A}_\textsc{{\tiny O}}&=\int p_\textsc{{\tiny O}}d\delta\theta, \quad
%\mathcal{B}_\textsc{{\tiny O}}=\int p_\textsc{{\tiny O}} e^{i\varphi_t}d\delta\theta, \quad p_\textsc{{\tiny o}}=2|t|^4|r|^4,
%\end{align*}
%\begin{align*}
%\mathcal{A}_\textsc{{\tiny O}}&=\int p_\textsc{{\tiny O}}d\delta\theta, \quad
%\mathcal{B}_\textsc{{\tiny O}}=\int p_\textsc{{\tiny O}} e^{i\varphi_t}d\delta\theta, \quad p_\textsc{{\tiny o}}=2|t|^4|r|^4,
%\end{align*}
The average here is taken over $\delta\theta =\theta-\theta_B$ since $\beta=\beta(\delta\theta)$ is a function of the angular deviation, and where $\theta_B$ is the Bragg angle.
The measured contrast and phase against $\delta\theta$ are shown in Fig.~\ref{Fig:3to4BNI}a and Fig.~\ref{Fig:3to4BNI}b, respectively. 
Also shown is the simulated momentum distribution $g(\delta\theta)$ accepted by a single crystal, where the full width at half maximum (FWHM) is given by Darwin width of the crystal $\sigma_\theta=4.26\ \mu\text{rad}$. 
%The distribution is a convolution of the added blade and the NI. Also shown is distribution with a calculated  FWHM of a perfect [111] Si crystal. 
The addition of an extra blade breaks the blade separation symmetry (equal separation between all the blades). The result of this is that the measured phase depicted on Fig.~\ref{Fig:3to4BNI}b is composed of the dynamical phase and the phase due to defocussing.  By separating these two phases, we extract a purely dynamical phase  given by Fig.\ref{Fig:DDPhaseD}. This is achieved by using the FWHM extracted from the experiments, and assuming that that momentum distribution only changes when the orientation of the crystal changes and not due to defocussing. A similar experiment has been done with the extra crystal blade oriented in the Bragg geometry \cite{Lemmel_2013_ActaCrys}.
%=================================================================================================================
%=================================================================================================================
By accounting for this dynamical phase we can estimate the maximum contrast of the four-blade NI. If the four-blade NI is made from 1 mm thick Si blades in the (111) crystallographic orientation (as per \cite{Pushin_2011_PRL}), and illuminated with neutrons of $\lambda = 2.71$ {\AA}, the estimated maximum contrast is $\sim 85\%$.
\subsection{Five-blade double loop neutron interferometer}

 The five-blade NI is similar to the four-blade NI but with an additional middle blade  $U_B$. With a phase $\phi$ in the first loop and $\chi$ in the second loop (see Fig.~\ref{Fig:345NI}c) the combined operation of the interferometer is:
\bea\nonumber\label{eqn:5bOperetor}
R_{5}&=&U_B R_x(\pi)U_B R_x(\pi) R_z(\phi) U_B,\\\nonumber
%&=&R_z(\beta)R_x(\alpha)R_z(\beta)R_z(\chi)R_x(\pi)
%\\\nonumber& &
%R_z(\beta)R_x(\alpha)R_z(\beta)R_x(\pi)R_z(\phi)R_z(\beta)R_x(\alpha)R_z(\beta),\\\nonumber
%&=&R_x(\alpha)R_z(\chi)R_x(\pi)R_x(\alpha)R_x(\pi)R_z(\phi)R_x(\alpha),\\
&=&R_z(\beta)R_x(\alpha)R_z(\chi)R_x(\alpha)R_z(\phi)R_x(\alpha)R_z(\beta).
%&=&R_x(3\alpha),
\eea
 With an incident state of $\ket{\text{I}}$ onto the NI, the intensity at the O- and H- beams is
\bea
I_{O5}(\phi,\chi)%&=&|\bra{\text{I}}R_{5}\ket{\text{I}}|^2\\
&=&\frac{1}{4}\Big[2+\cos(\chi-\phi)-\cos(\chi+\phi)\Big],\\
I_{H5}(\phi,\chi)%&=&|\bra{\text{II}}R_{5}\ket{\text{II}}^2\\
&=&\frac{1}{4}\Big[2-\cos(\chi-\phi)+\cos(\chi+\phi)\Big].
\eea
Notice that, there is no dependence on $\beta$ and so the dynamical phase is refocused.
The refocusing of the dynamical phase can also be understood in the sense of chirping, as the wavevectors that were travelling faster than the mean wavevector before the second blade (which acts as a mirror) tend to travel slower than the mean wavevector after the mirror blade (and vice-versa). 
This is the same principle of noise refocusing which is employed in nuclear magnetic resonance \cite{Hahn_1950_PhysRev, Becker_1959_JAMChem, Carr_1954_PhysRev}.

%\subsubsection{Effects five  blade  (Sagnac) NI}
%This interferometer is similar to a double loop $MZ$ interferometer but the middle blade is offset from the center, and hence the two neutron beam paths do not interfere in the middle of the interferometer (figure 1d). Therefore this middle blade now acts only as a $R_z(\beta)$. The sequence becomes:
%\bea\nonumber
%R_{S}^5&=&\overbrace{R_z(\beta)R_x(\alpha) R_z(\beta)}\overbrace{R_x(\pi)}
%\overbrace{R_z(\beta)R_x(2\pi)R_z(\beta)}\\\nonumber & &
%\overbrace{R_x(\pi)}R_z(\phi)\overbrace{R_z(\beta)R_x(\alpha)R_z(\beta)},\\
%&=&R_x(\alpha)R_x(\pi) R_x(\pi)R_z(\phi)R_x(\alpha)
%%&=&R_x(2\alpha),
%\eea
%

We conclude that the three-blade and five-blade NIs are immune to dynamical phase noise originating from the momentum spread of the incoming neutrons, while the four-blade NI is not. 
%This implies that as the momentum spread of the incoming beam increases the contrast of the four blade decreases. 
Next, we will analyse and compare the performance of these interferometers against external vibrational noise.

%============================================================
%Sec: Mechanical Vibration
\section{Effects of Mechanical Vibration}
\label{Sec:MechVib}
The effect of mechanical vibrations on matter wave interferometry has been studied for specific implementations \cite{Bongs_2001_PRA}. In neutron interferometry mechanical vibrations are commonly reduced by using vibration isolation systems although, the effect of low-frequency vibration still persists. 
The four-blade NI has the experimentally demonstrated advantage  over the three-blade NI of being robust against slow varying external mechanical vibrational noise \cite{Pushin_2011_PRL}. 
In this article we adopt the vibration model in \cite{Pushin_2009_PRA}, which treats vibrations as sinusoidal oscillations in the form  $\zeta(t)=\zeta_0\sin(\omega t +\varphi)$ where $\zeta_0$ is the amplitude of the noise, $\omega$ is the frequency and  $\varphi\in[0,2\pi]$ is a random phase that considers different arrival times of the neutrons at the first blade. 
Mechanical vibrations may change the momentum of the neutron which leads to a phase difference around any closed interferometry loop
\bea\nonumber
\Delta\Phi
&=&\frac{1}{\hbar}\int_\text{path I} \vec{p}_\text{I}\cdot d\vec{s}-\frac{1}{\hbar}\int_\text{path II} \vec{p}_\text{II}\cdot d\vec{s}
=\frac{1}{\hbar}\oint \Delta\vec{p}\cdot d\vec{s},
\eea
where $\vec{p}_\text{I}$ and $\vec{p}_\text{II}$ are associated momentum changes for path I and path II, respectively, and $\Delta\vec{p}=\vec{p}_\text{I}-\vec{p}_\text{II}$. The main contributions to the decrease in coherence comes from the translational vibration noise along the $y$-axis (y-noise) and rotation vibration around the $z$-axis ($\theta_z$ referred to simply as z-noise). The y-noise comes from the interferometer vibrations along the reciprocal lattice vector, and the z-noise from rotations around the axis perpendicular to plane of interference. Using the form of the noise stated above, the y-noise is  modelled as  $y(t)=y_0\sin(\omega t +\varphi)$, and the z-noise as $\theta(t)=\theta_0\sin(\omega t +\varphi)$. The frequency of the noise along the y-axis and the z-axis is not necessarily the same. 

%In our simulations the rotation and translation noise amplitudes is chosen to be 10$^{-6}$, the wavelength is 4.4 {\AA} (which is the wavelength for cool neutron interferometry source at the NIST Center for Neutron Research), and the separation between the first and second blade of the interferometer is 5 cm (which roughly corresponds to the actual interferometer).

%
%%%%%=================================================================================================================
%\begin{figure}
%\center
%\includegraphics[scale=0.3]{5BladeNI.pdf} 
%\caption{a)Sketch of the five blade double MZ loop NI.}\label{fig:5BladeNI}
%\end{figure}
%%%%=================================================================================================================

\subsection{Y-noise}

\begin{figure}[!t]
\center
\includegraphics[width=.95\columnwidth]{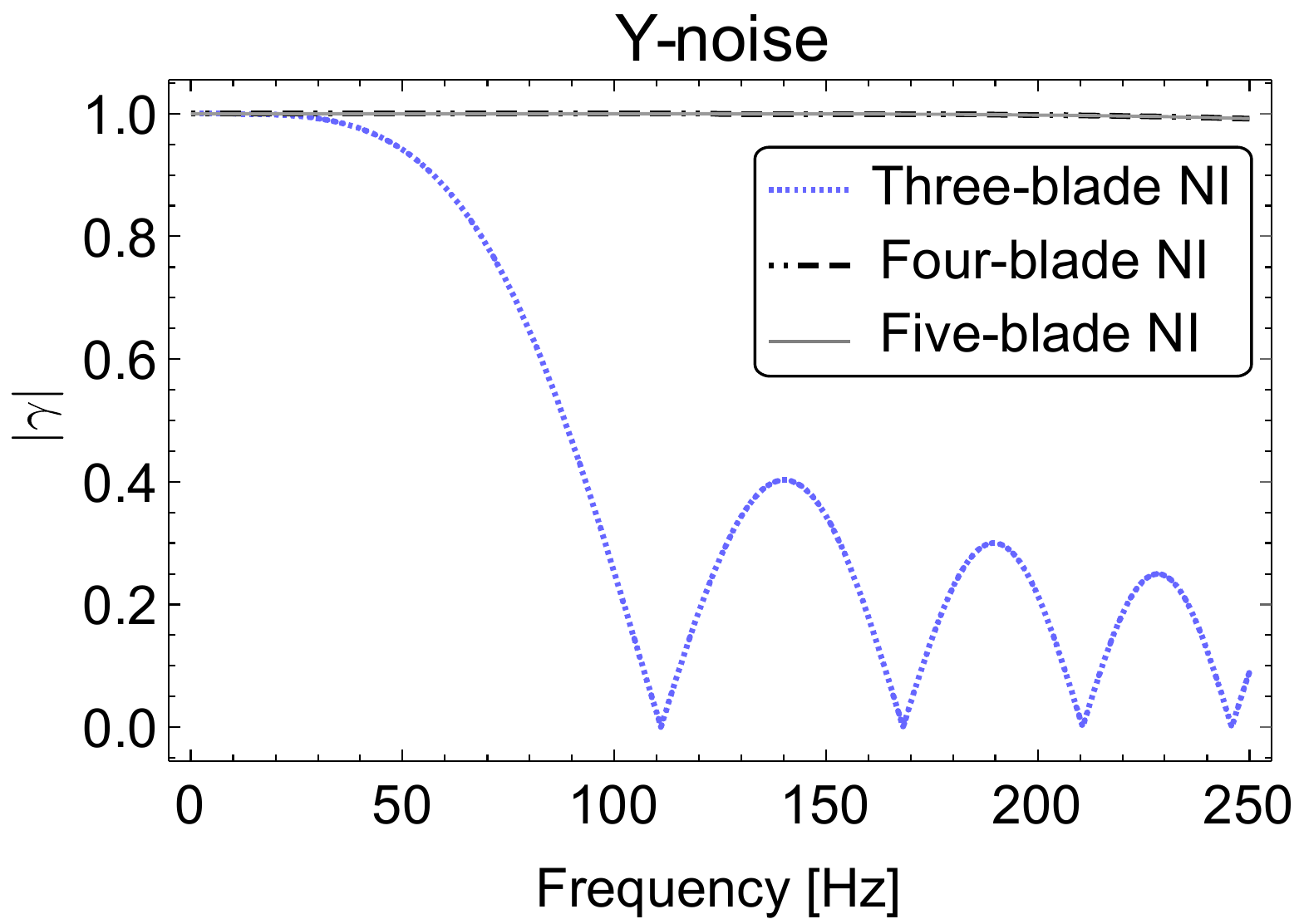}
\caption{Simulations of the variation of the absolute value of the coherence function $\gamma$ versus strength $\omega$ for each of three, four and five blade neutron interferometers. The coherence of the four and five blade NI is not affected for low frequencies although they are affected at  freqiencies of $\omega>250$ Hz. Note that the decoherence free condition from the configuration where $\phi=-\chi+\pi$ is used. }\label{Fig:YNoise}
\end{figure}

Let the velocity of the incidence neutron be decomposed into two components, perpendicular and parallel to the reciprocal lattice vector $\mathbf{ v}=v_\perp \hat{e}_x+v_\parallel\hat{e}_y$. If the interaction of the neutron with the blade is modelled as a bouncing ball from a hard surface,  the velocity along the $x$-axis is not affected while that along the $y$-axis is $v_y=-v_\parallel+2{u}_y(t)$, where ${u}_y(t)=d{y}(t)/dt$ is the time derivative. 
Assume that the neutron enters the interferometer at $t=0$, the phase shift between path I and path II caused by y-noise vibrations in a three blade NI is:
\bea
\Delta\Phi(\varphi)
&=&\frac{32m}{\hbar}\tau^2[v_\parallel-{u}_y(0)]\dot{u}_y(0),
\eea
where $m$ is the mass of the neutron, $\tau=L/v_\perp$ and $L$ is the distance between the first and second blades of the interferometer. 
For low frequency noise where $\omega\tau\ll 1$:
\bea
\Delta\Phi(\varphi)=\frac{32mv_yy_0\tau^2}{\hbar}\omega^2\sin\varphi,
\eea
since  $v_y\gg{u}_y(0)$.
The probability of detecting a single neutron at the O- and H- detectors in the three blade NI is
\bea
I_{O3}(\phi)&=&\frac{1}{2}\Big[1+\cos\left(\phi+\Delta\Phi(\varphi)\right)\Big],\\
I_{H3}(\phi)&=&\frac{1}{2}\Big[1-\cos\left(\phi+\Delta\Phi(\varphi)\right)\Big].
\eea
Each neutron arrives at the first blade at different instances and picks a different initial phase $\varphi$. Integrating over a uniform probability distribution  $p(\varphi)=1/2\pi$, the average intensity at the O-beam is
\begin{align}
\overline{I_{O3}}(\phi)
&=\frac{1}{2}\Big[1+|\gamma|\cos(\phi +\text{arg}[\gamma])\Big],
\label{Eqn:IAve3}
\end{align}
where $\gamma$ is the coherence function which is defined as for statistically stable  noise \cite{Petrascheck_1984_ActaCrys,Ewolf_2007_Book}
\begin{align}
\gamma&=\frac{1}{2\pi}\int_0^{2\pi} \exp[i \Delta\Phi(\varphi)] d\varphi.\label{Eqn:cohFunc3BNI}
\end{align}
The absolute value of coherence function, $|\gamma|$, for the three-blade NI is equal to the contrast $\mathcal{V}$ defined in Eq.~(\ref{Eq:Contrast}). 
 We consider an interferometer with $L=5$ cm, a wavelength of 4.4 {\AA}, and a y-noise with an amplitude of $y_0=0.1$ $\mu$m. Using these values the coherence function for the y-noise in the three blade NI reduces to
\begin{align}
\gamma&=J_0(\Omega \omega^2),\quad \text{ with }\quad \Omega= \frac{32mv_yy_0\tau^2}{\hbar}
\label{Eq:Coh3y}
\end{align}
where $J_0$ is the Bessel function of the first kind.
Shown in Fig.~\ref{Fig:YNoise} is $\mathcal{V}=|\gamma|$  vs the noise strength $\omega$, where it can be compared to the four-blade and five-blade NIs.

%%%=================================================================================================================
%%%=================================================================================================================
%

In the four-blade NI the phase difference in the first loop $\Delta\Phi_{1}$, and the phase difference in the second loop $\Delta\Phi_{2}$ (see \ref{Fig:345NI}b for the loop labels) are
\bea
\Delta\Phi_{1}&=&-\frac{4m\tau^2}{\hbar}[v_\parallel-u_y(0)][2\dot{u}_y(0)+\tau\ddot{u}_y(0)],\\
\Delta\Phi_{2}&=&\frac{4m\tau^2}{\hbar}[v_\parallel-u_{y}(0)][2\dot{u}_y(0)+7\tau\ddot{u}_y(0)].
\eea
The phase difference is effectively the sum of the phases in loops 1 and 2, and for low frequency noise where $\omega\tau\ll 1$,  the phase difference is given by
%\bea\nonumber
%\Delta\Phi_{2}=-\Delta\Phi_{1}= \frac{8m\tau^2}{\hbar}v_y\dot{u}_y(0),
%\eea
\bea
\Delta\Phi(\varphi)
 =\frac{24mv_yy_0\tau^3}{\hbar}\omega^3\cos\varphi.\label{eqn:4Bphase}
\eea
 The probability of detecting a single neutron at the O- and H- detectors in the four blade NI is
\bea
I_{O4}(\phi)&=&\frac{1}{2}\Big[1-\cos(\phi+\Delta\Phi(\varphi))\Big],\\
I_{H4}(\phi)&=&\frac{1}{2}\Big[1+\cos(\phi+\Delta\Phi(\varphi))\Big].
\eea
Taking the average over the uniform phase distribution of $\varphi$, and considering the H-beam in the DFS as it carries the same phase information as the O-beam in the three-blade NI, the intensity of the DFS is
\begin{align}
\overline{I_{H4}}(\phi)
%&=\frac{1}{4\pi}\int_0^{2\pi}\Big(1+ \cos[\phi+\Delta\Phi(\varphi)]\Big) d\varphi,\\
&=\frac{1}{2}\Big(1+\gamma \cos\phi\Big).
\label{Eqn:IAve4}
\end{align}
Where, the coherence similar to the one for the three blade NI is
\begin{align}
\gamma&=J_0(\Omega \omega^3),\quad \text{ with }\quad \Omega= \frac{24mv_yy_0\tau^3}{\hbar},
\end{align}
The coherence function $|\gamma|=\mathcal{V}$ for the four blade NI under the influence of y-noise is compared to the three blade and five blade in Fig.~\ref{Fig:YNoise}.

%=================================================================================================================
\begin{figure}[t!]
\center
\includegraphics[width=0.95\columnwidth]{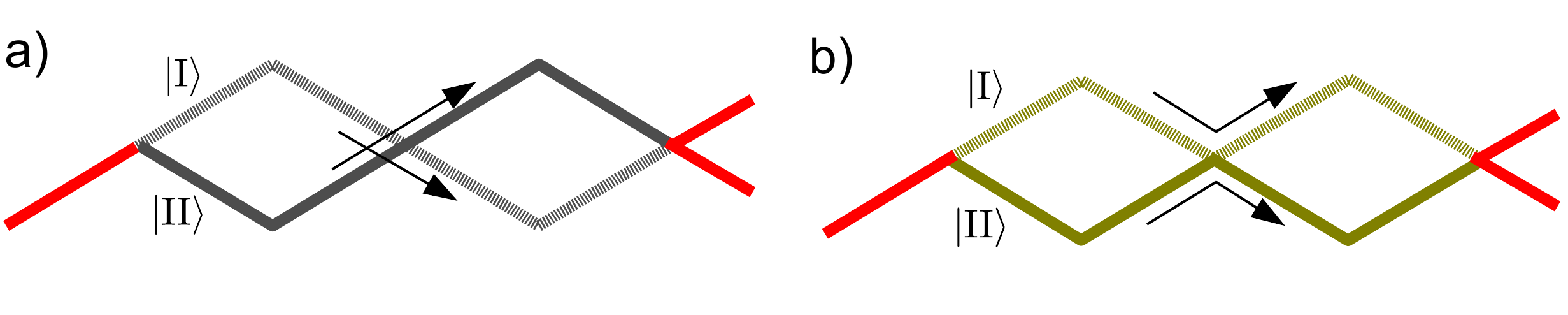} 
\caption{The five blade interfrometer split into to four paths which constitute two cases. In (a) the symmetric case  $\ket{\text{I}}\rightarrow\ket{\text{I}}$   and $\ket{\text{II}}\rightarrow\ket{\text{II}}$. In  (b) the anti-symmetric case $\ket{\text{I}}\rightarrow\ket{\text{II}}$   and $\ket{\text{II}}\rightarrow\ket{\text{I}}$. }\label{Fig:5ASNI}
\end{figure}
%=================================================================================================================

In the five blade NI we first resolve the path taken by the neutron to the last blade into four trajectories. For simplicity, we split the four trajectories into two categories, the $\textit{symmetric}$ case and the $\textit{antisymmetric}$ case. The symmetric case contains the two paths corresponding to the middle blade acting as a perfect transmitter ($\ket{\text{I}}\rightarrow\ket{\text{I}}$   and $\ket{\text{II}}\rightarrow\ket{\text{II}}$ , see Fig.~\ref{Fig:5ASNI}a, and the antisymmetric case is where the middle blade acts as a perfect reflector ($\ket{\text{I}}\rightarrow\ket{\text{II}}$   and $\ket{\text{II}}\rightarrow\ket{\text{I}}$ , see Fig.~\ref{Fig:5ASNI}b). 
The symmetric case is identical to the four blade NI. In a similar way, we split the total phase into two  components. 
In the symmetric case the phases denoted by $\Delta\Phi_{1}$ and $\Delta\Phi_{2}$ for loop 1 and loop 2 respectively are
\bea 
\Delta\Phi_{1}&=&-\frac{4m\tau^2}{\hbar}[v_\parallel-u_y(0)][2\dot{u}_y(0)+\tau\ddot{u}_y(0)],\\
\Delta\Phi_{2}&=&\frac{4m\tau^2}{\hbar}[v_\parallel-u_y(0)][2\dot{u}_y(0)+7\tau\ddot{u}_y(0)],\label{eqn:synoise}
\eea
In the antisymmetric case the phases in loop 1 and 2 denoted by $\Delta\Phi^{\prime}_{1}$ and $\Delta\Phi^{\prime}_{2}$ respectively are
\bea
\Delta\Phi_{1}^{\prime}&=&\Delta\Phi_{1},\\
\Delta\Phi_{2}^{\prime}&=-&\frac{4m\tau^2}{\hbar}[v_\parallel-u_y(0)][2u_y(0)+3\tau\ddot{u}_y(0)].\label{eqn:asynoise}
\eea
In the low frequency noise regime where $\tau\omega\ll1$
%\bea\label{eqn:5Bphase}
%\Delta\Phi_{1}^{\prime}=\Delta\Phi_{2}^{\prime} =\Delta\Phi_{1}=
%-\Delta\Phi_{2}\approx -\frac{8m\tau^2}{\hbar}v_y\dot{u}_y(0).
%\eea
%In a five blade NI, 
the resulting phase difference in the symmetric case and the antisymmetric case is
%\bea\nonumber
%\Delta\Phi(\varphi)&=& \Delta\Phi_{1}+\Delta\Phi_{2}\\ \label{Eq:symPhasey} &=& \frac{24mv_yy_0\tau^3}{\hbar}\omega^3\cos\varphi, \quad \text{ symmetric}\\\nonumber
%\Delta\Phi^{\prime}(\varphi)&=& \Delta\Phi_{1}^\prime+\Delta\Phi_{2}^\prime\\  &=& \frac{16mv_yy_0\tau^2}{\hbar}\omega^2\sin\varphi, \quad \text{antisymmetric}.\label{Eq:antisymPhasey}
%\eea 
\bea\label{Eq:symPhasey}
\Delta\Phi(\varphi)  &=& \frac{24mv_\parallel y_0\tau^3}{\hbar}\omega^3\cos\varphi, \quad \text{ symmetric}\\
\Delta\Phi^{\prime}(\varphi) &=& \frac{16mv_\parallel y_0\tau^2}{\hbar}\omega^2\sin\varphi, \quad \text{antisymmetric}.\label{Eq:antisymPhasey}
\eea 
The phase difference from external vibrations along the y-axis cancel out in the symmetric case, but effectively doubles in the anti-symmetric loop. The effect of this noise and conditions under which it can be removed will be discussed later. Prior to that we examine the effect of y-noise.

With a phase $\phi$ in the first loop and $\chi$ in the second loop (see Fig.~\ref{Fig:345NI}c), the probability of detecting a single neutron at the O- and H- detectors of the five-blade NI are
\begin{align}\nonumber
I_{O5}(\phi,\chi)=\frac{1}{4}\Big(2&+\cos[\chi-\phi+\Delta\Phi(\varphi)]\\ &-\cos[\chi+\phi+\Delta\Phi^\prime(\varphi)]\Big),\\\nonumber
I_{H5}(\phi,\chi)=\frac{1}{4}\Big(2&-\cos[\chi-\phi+\Delta\Phi(\varphi)]\\ &+\cos[\chi+\phi+\Delta\Phi^\prime(\varphi)]\Big).
\end{align}
where, the symmetric $\Delta\Phi(\varphi)$ and the antisymmetric phase differences are defined in Eqs.~(\ref{Eq:symPhasey}) and (\ref{Eq:antisymPhasey}).
The average H-beam intensity over the uniform distribution of $\varphi$ of the H-beam is
\begin{align}\nonumber
\overline{I_{H5}}(\phi,\chi)
%&=\frac{1}{4}\Big(2-\gamma_1\cos\left(\chi-\phi\right)- \gamma_2\sin\left(\chi-\phi\right)\\\nonumber &+\gamma_3\cos\left(\chi+\phi\right)
% +\gamma_4\sin\left(\chi+\phi\right) \Big),\\
 &=\frac{1}{4}\Big(2-|\gamma| \cos(\chi-\phi+\arg\gamma ) \\
 &+|\gamma^\prime|\cos(\chi+\phi+\arg\gamma^\prime)\Big),
 \label{Eqn:IAvey5}
\end{align}
%\bea
%\overline{I_{H5}(\phi,\chi)}&=&\frac{1}{4}\Big(2+\mathcal{V}_1\cos\left(\chi+\phi\right) +\mathcal{V}_3\cos\left(\chi-\phi\right)
%+\mathcal{V}_2\sin\left(\chi+\phi\right) +\mathcal{V}_4\sin\left(\chi-\phi\right) \Big).\label{eqn:IH5form2}
%\eea
%$\gamma_1,\gamma_2$ are the real and imaginary parts of $\gamma$ respectively, and  $\gamma_3,\gamma_4$ are the real and the imaginary parts of $\gamma^\prime$ 
With the coherence function of the symmetric and anti-symmetric cases:
\begin{align}\nonumber
\gamma&=\frac{1}{2\pi}\int_0^{2\pi} \exp[i \Delta\Phi(\varphi)] d\varphi, \\
\gamma^\prime&=\frac{1}{2\pi}\int_0^{2\pi} \exp[i \Delta\Phi^\prime(\varphi)] d\varphi.
\end{align}
%The $\gamma_i$s define the normalized amplitude of each of the contributing oscillations.
For y-noise, it can be shown that,
\bea
\gamma&=&J_0(\Omega \omega^3),\quad \text{ with }\quad \Omega=\frac{24mv_\parallel y_0\tau^3}{\hbar},\\
\gamma^\prime&=&J_0(\Omega^\prime \omega^2),\quad \text{ with }\quad \Omega^\prime=\frac{16mv_\parallel y_0\tau^2}{\hbar}.
\eea

\begin{figure}[!t]
\center
\includegraphics[width=\columnwidth]{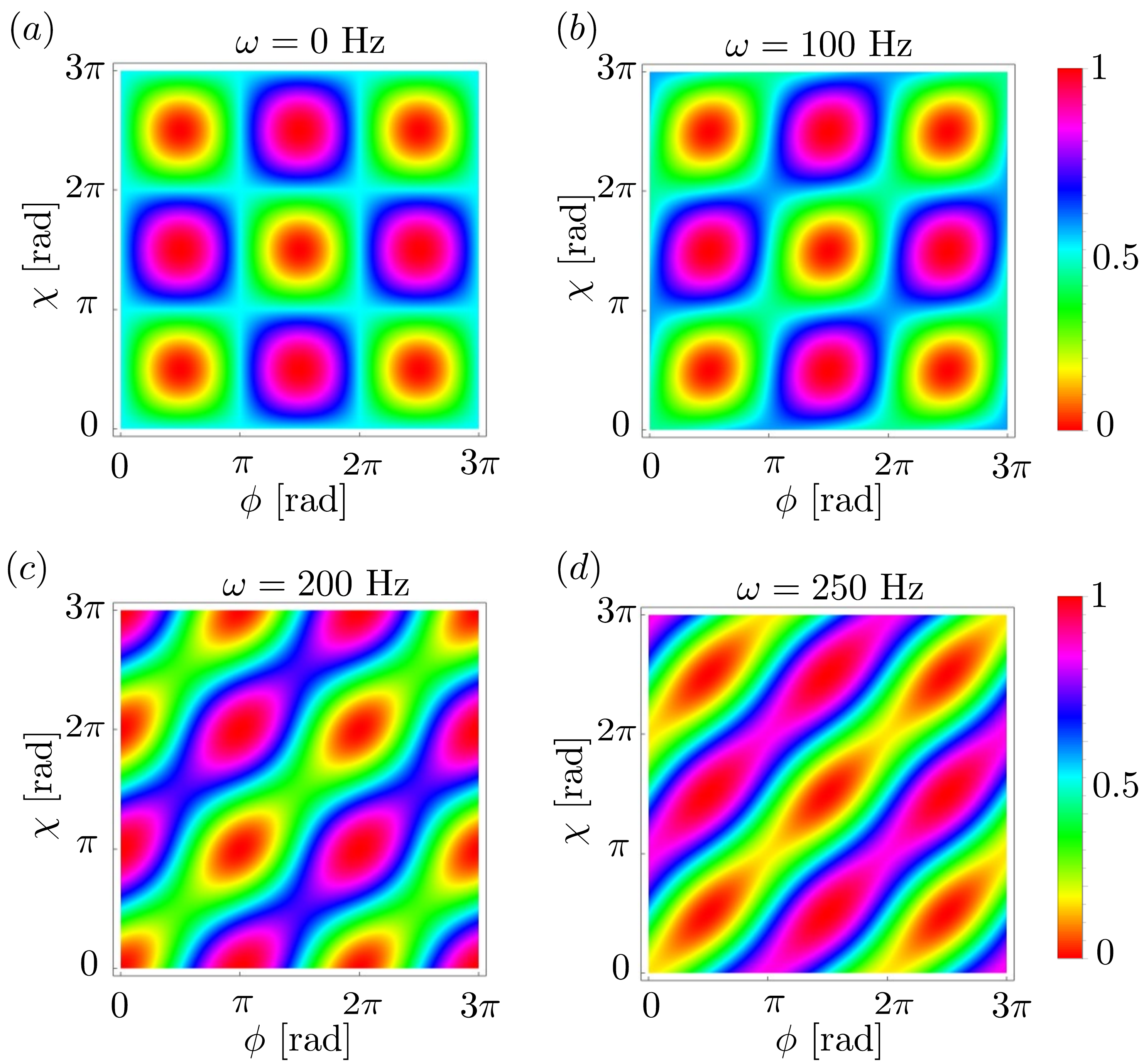} 
\caption{ Density plot of the intensity at the H-beam for a five blade-neutron neutron interferometer as a function of the phase, $\phi$, in loop 1, and phase, $\chi$, in loop 2. (a) The intensity without noise. In this plot the oscillations are clearly visible along any line in the 2D plane ($\phi,\chi$). The intensity in the presence of y-noise is shown in (b)  with $\omega=100$ Hz, in (c)  with $\omega=200$ Hz, and in (d) for $\omega=250$ Hz. The interference pattern is dampened along some configurations of $\phi$ and $\chi$ for example $\chi=\phi + $constant. In the simulation, the interferometer separation between blades $L=5$ cm, the neutron wavelength is 4.4 {\AA}. }\label{Fig:DensityY}
\end{figure}

Consider an interferometer where the amplitude of y-noise is $y_0=0.1$ $\mu$m. The H-beam intensity without noise ($\omega=0$) is presented on Fig.~\ref{Fig:DensityY}a. 
In Figs.~\ref{Fig:DensityY}b, \ref{Fig:DensityY}c, and \ref{Fig:DensityY}d the same intensity is plotted for y-noise with $\omega=150$ Hz, $\omega=200$ Hz and $\omega=250$, respectively.
The region through the density plots where the oscillations are dampened illustrates the effect of noise. It is clearly visible on the plot that there are some combinations of the phase on the first and second loop for which the effect of noise is minimal. 
One obvious choice from Fig.~\ref{Fig:DensityY}b is the vertical line $\phi=\pi$, however, this line is only unique for $\omega=200$ Hz. For a different $\omega$ a different vertical line would be required. On the other hand, the set of conditions which include the lines $\phi=-\chi+\mu$, where $\mu$ is a constant, is capable of refocusing  any low-frequency mechanical vibrational noise. Along these lines, the effect of noise results in a DC shift of the intensity profile with no effect on coherence. 

%%%%=================================================================================================================
%%%=================================================================================================================

\begin{figure}[!t]
\center
\includegraphics[width=.95\columnwidth]{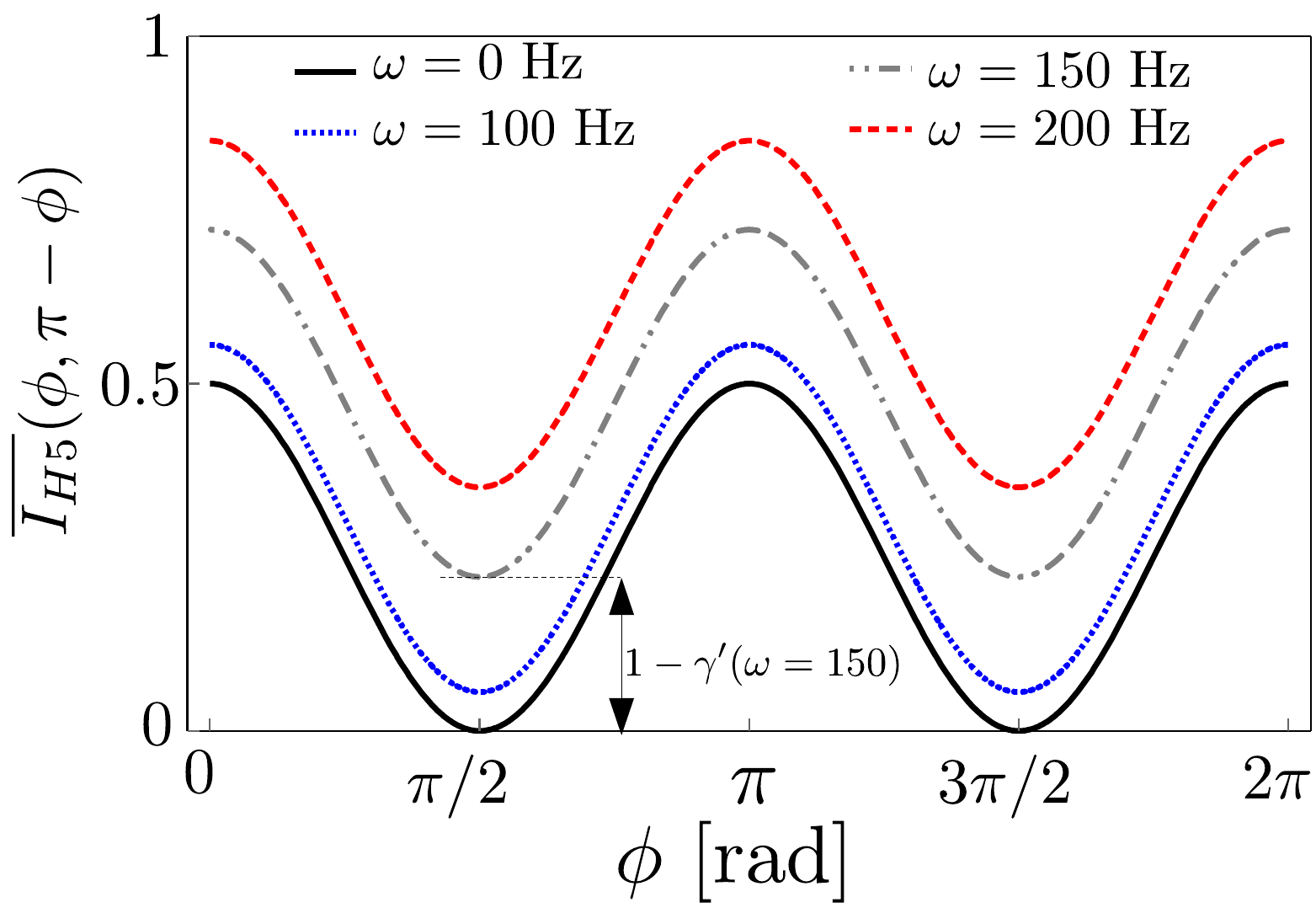} 
\caption{ The H-beam intensity as a function of the correlated phase $\phi$ for low-frequency y-noise with $\omega=0,100,150,$ and  200 Hz. When the noise refocusing condition $\phi+\chi=\pi$ is used, the effect of noise is simply an additional background term, with a magnitude of $1-\gamma^\prime$. }\label{Fig:IntenHNN}
\end{figure}

By choosing $\mu=\pi$ to get $\chi=\pi-\phi$, the intensity in the presence of y-noise can be expressed as 
%$\gamma$ is real and $\gamma^\prime=0$ leading to
%\bea
%\overline{I_{H5}(\phi,\chi)}&=&\frac{1}{4}\Big[2+\gamma_4+\gamma_1\cos\left(2\phi-\pi/2\right) +\gamma_2\sin\left(2\phi-\pi/2\right) \Big],\\
%&=&\frac{1}{4}\Big[2+\gamma_4+|\gamma|\sin\left(2\phi -\arg\gamma\right) \Big]
%\label{eqn:IH5formF}
%\eea
\begin{align}
\overline{I_{H5}}(\phi,\pi-\phi)
%&=\frac{1}{4}\Big(2+\gamma_4-\gamma_1\cos\left(2\phi-\pi/2\right) \\&-\gamma_2\sin\left(2\phi-\pi/2\right) \Big),\\
&=\frac{1}{4}\Big[2-\gamma^\prime-|\gamma|\sin\left(2\phi+\arg(\gamma) \right) \Big].
\label{eqn:IH5formF}
\end{align}
In the five-blade NI noise acts as a DC shift, or an additional background contribution of $1-\gamma^\prime$. 
This is shown in Fig.~\ref{Fig:IntenHNN}. As the noise increases, the interference pattern is displaced along the vertical axis.
Even though, the coherence, or the depth of the modulation, $|\gamma|$ remains the same, the contrast as defined by Eq.~(\ref{Eq:Contrast}) reduces. For a y-noise of 100 Hz, the interferogram is offset by 0.2 which results in a relative contrast of about 82\%.
In Fig.~\ref{Fig:YNoise} a plot of $|\gamma|$ for the five-blade NI is compared with that for the three and four blade NIs.
Therefore, the five-blade NI is capable of refocusing low-frequency noise just as the four-blade DFS NI.
The coherence of the four-blade and five-blade NIs is not noticeably affected at low frequencies, although they start to get affected at frequencies above 250 Hz.
%We note here that  since the intensity is not balanced, the absolute value of the coherence function is not equal to the contrast.  Under this specific noise refocusing condition, the contrast of the five-blade neutron interferometer obtained using Eq.~(\ref{Eq:Contrast}) is $\gamma/(2-\gamma^\prime)$.
%%%%=================================================================================================================
%%%=================================================================================================================

\subsection{Z-noise}

The noise around the $z$-axis is modeled as $\theta(t)=\theta_0\sin(\omega t +\varphi)$. Again assuming  an incident neutron on the first blade at $t=0$ and using small angle approximations the phase difference for a three blade NI is
\bea
\Delta\Phi(\varphi)&=&\frac{32m\tau}{\hbar}[v_\parallel-2L\dot{\theta}(0)]L\dot{\theta}(0),\\
&=&\frac{32mv_\perp v_\parallel\theta_0\tau^2}{\hbar}\omega\cos\varphi.
%&=&\frac{32mL\tau}{\hbar}v_y\dot{\theta}(0)
%\Delta\Phi^3_{MZ} (\varphi)&=&\frac{32m}{\hbar}\tau L\omega\theta_0\sin\varphi(2L\omega\theta_0\sin\varphi-v_y)
\eea
%Using this, a coherence function similar to Eq.~\ref{Eq:Coh3y} can be define.
 In the four-blade NI the phase difference, $\Delta\Phi_{1}$, in loops 1  and, $\Delta\Phi_{2}$, in loop 2  are given by
\bea
\Delta\Phi_{1}&=&\frac{8m\tau}{\hbar}[v_\parallel-2L\dot{\theta}(0)][L\dot{\theta}(0)-L\tau\ddot{\theta}(0)],\\
\Delta\Phi_{2}&=&-\frac{8m\tau}{\hbar}[v_\parallel-2L\dot{\theta}(0)][L\dot{\theta}(0)+5L\tau\ddot{\theta}(0)],
\eea
such that the low frequency phase difference is
\begin{align}
\Delta\Phi
= -\frac{48mv_\perp v_\parallel\theta_0\tau^3}{\hbar}\omega^2\sin\varphi.
\end{align}
%\bea
%\Delta\Phi_{SAG}^4(\varphi)&=&=\frac{48m\tau^3}{\hbar}L\omega^2\theta_0\cos\varphi(2L\omega\theta_0\sin\varphi-v_y)
%\eea
%With this phase the intensity at the O-- and H--beams can be derived.
%The four blade NI is immune to noise due to external vibrations. 

The phase difference for the five-blade NI is again split into two components. The symmetric phase difference acquired in loop one $\Delta\Phi_{1}$ and loop two $\Delta\Phi_{2}$ are
\bea
\Delta\Phi_{1}&=&\frac{8m\tau}{\hbar}[v_\parallel-2L\dot{\theta}(0)][L\dot{\theta}(0)-L\tau\ddot{\theta}(0)],\\
\Delta\Phi_{2}&=&-\frac{8m\tau}{\hbar}[v_\parallel-2L\dot{\theta}(0)][L\dot{\theta}(0)+5L\tau\ddot{\theta}(0)],\label{eqn:sznoise}
\eea
and the phase of loop 1 and loop 2 in the antisymmetric case are,
\bea
\Delta\Phi_{1}^{\prime}&=&\Delta\Phi_{1},\\
\Delta\Phi_{2}^{\prime}&=&\frac{8m\tau}{\hbar}[v_\parallel-2L\dot{\theta}(0)][L\dot{\theta}(0)+L\tau\ddot{\theta}(0)],\label{eqn:asznoise}
\eea
For low frequency noise where $\tau\omega\ll1$,
%\bea\nonumber
%\Delta\Phi_{1}^{\prime}&=&\Delta\Phi_{2}^{\prime} =\Delta\Phi_{1}
%=-\Delta\Phi_{2}= \frac{8mLv_y\theta_0\tau}{\hbar}\omega\cos\varphi,
%\eea 
the phase difference in the symmetric case and the antisymmetric case are
%\begin{align}
%\Delta\Phi(\varphi)&= \Delta\Phi_{1}+\Delta\Phi_{2} = -\frac{48mLv_y\theta_0\tau^2}{\hbar}\omega^2\sin\varphi, \label{Eq:symPhasez}\\
%\Delta\Phi^{\prime}(\varphi)&= \Delta\Phi_{1}^\prime+\Delta\Phi_{2}^\prime= \frac{16mLv_y\theta_0\tau}{\hbar}\omega\sin\varphi.\label{Eq:antisymPhasez}
%\end{align}
\begin{align}
\Delta\Phi(\varphi)&= -\frac{48mv_\perp v_\parallel\theta_0\tau^3}{\hbar}\omega^2\sin\varphi, \quad \text{ symmetric} \label{Eq:symPhasez}\\
\Delta\Phi^{\prime}(\varphi)&= \frac{16mv_\perp v_\parallel\theta_0\tau^2}{\hbar}\omega\sin\varphi, \ \text{ antisymmetric}.\label{Eq:antisymPhasez}
\end{align}
Just like the y-noise, the phase difference from external vibrations along the z-axis cancel out in the symmetric, but effectively double in the anti-symmetric case. 
The effect of this noise and conditions under which it can be removed are similar to the y-noise.

%%%=================================================================================================================
%
%%%%%%%%%%%%%%%%%%%%%%%%Contrast FIG here%%%%%%%%%%%%%%%%%%
%%%=================================================================================================================
\begin{figure}[!t]
\center
\includegraphics[width=.95\columnwidth]{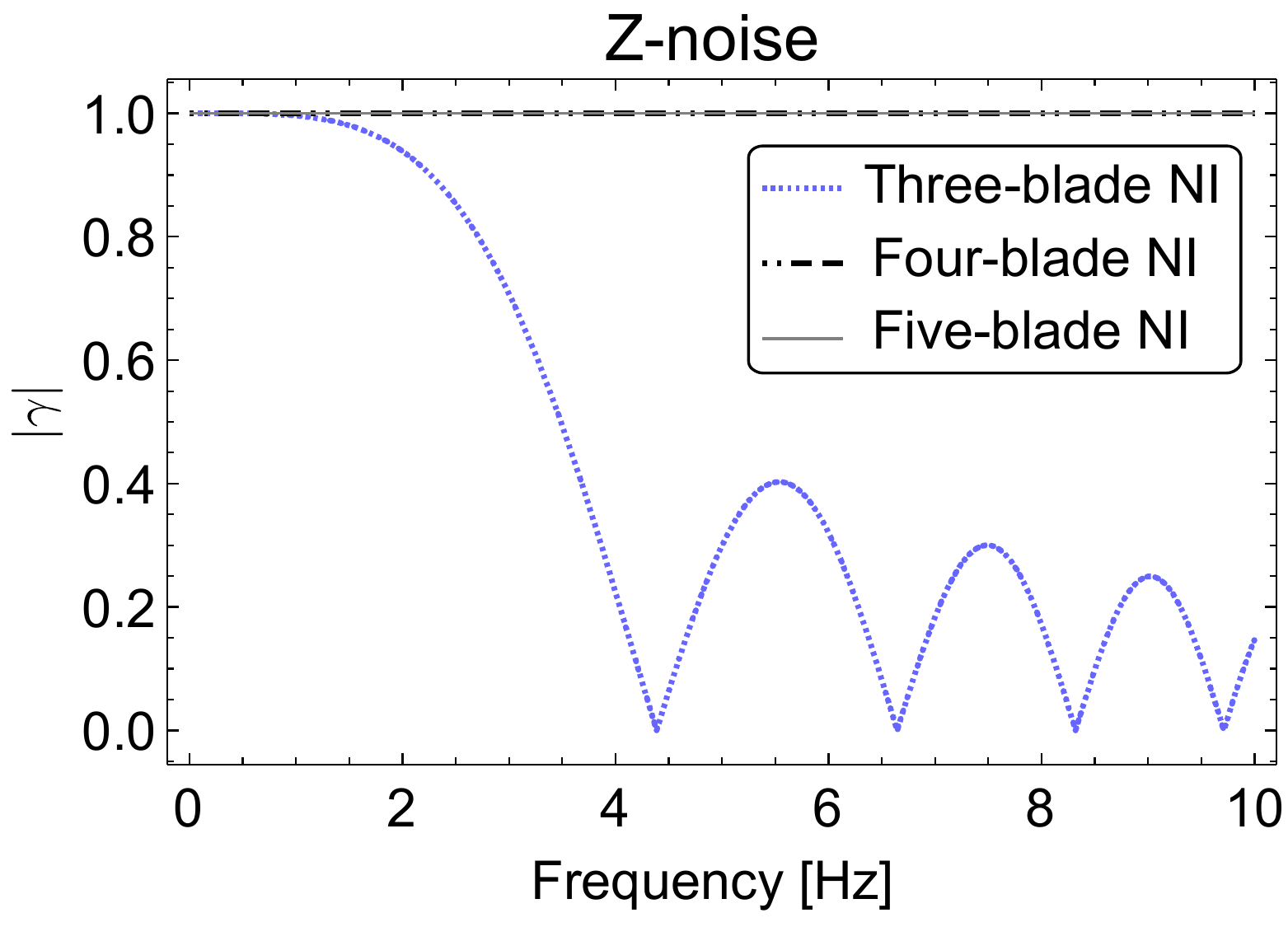}
\caption{Simulations of the variation of the absolute value of the coherence function with z-noise strength for each of the various neutron interferometers. The coherence function of the four blade and five blade interferometers remain unchanged at frequencies greater than 4 Hz while  the three blade NI is significantly affected. Note that the decoherence free condition from the configuration of the phase flags $\phi=-\chi+\pi$ is used.}\label{Fig:ZNoise}
\end{figure}
%%%=================================================================================================================

The coherence function can be calculated for the z-noise just as was done for the y-noise. 
In Fig.~\ref{Fig:ZNoise} is the absolute value of the coherence function $\gamma$ with frequency $\omega$ for vibrations around the z-axis. The vibration amplitude is $\theta_0=$ 0.1 $\mu$rads, with other conditions maintained as for the y-noise. 
The coherence function of the four-blade and five-blade interferometers remain unchanged at higher frequencies where  the three-blade NI is significantly affected for noise with frequencies greater than 4 Hz. 

It is worth noting here that the noise refocusing strength of the five-blade NI goes beyond the symmetric noise that is refocused by the four blade neutron interferometer. If the noise is antisymmetric, the five blade NI still retains the  ability to refocus but with the configuration changed to $\phi=\chi+\mu$. The four blade DFS NI does not have the ability to refocus this class of noise.

\section{Conclusion}

We have used a recently published approach in studying the effects of dynamical phase noise. We showed that this noise is refocused in a proposed five blade neutron interferometer, which is also insensitive to both dynamical and low frequency vibration noise. 
The power of the five blade neutron interferometer includes that it can also refocus antisymmetric noise. This class of noise could originate from various gradients (i.e. magnetic, temperature). 
From the analyses, we have a theory that can be generalized to any interferometer geometry to understand noise effects. The concepts presented here can be adapted to other matter-wave interferometers. Similar quantities related to the coherence can be extracted from various quantum systems in order to characterize noise \cite{Pushin_2007_Thesis}.  Our future plan is to test these concepts experimentally.
%In Ramsey--type interferometry for spin systems pulse sequences are used to refocus noise from magnetic fields  \cite{HahnE,Carrpurcell}. Techniques analogous to this are used in neutron--spin echo interference to refocus the effect of momentum distribution \cite{NSE}.

\section{Acknowledgements}
This work was supported by the Canadian Excellence Research Chairs (CERC) program (215284), the Natural Sciences and Engineering Research Council of Canada (NSERC) Discovery program, Collaborative Research and Training Experience (CREATE) program and the Canada First Research Excellence Fund (CFREF). M. Huber would like to thank Fred E. Wietfeldt for useful discussions and to appreciate the support of the National Science Foundation (NSF PHY-0245679).

%==============================================================
%:References
%==============================================================
%\bibliographystyle{unsrt}
\bibliographystyle{apsrev4-1}

\bibliography{QIFiveBladeNI}

\end{document}